\documentclass[aps,prl,a4paper,preprintnumbers,twocolumn,floatfix,showpacs]{revtex4}
\usepackage{amsmath,amsfonts,amssymb}
\usepackage{graphicx}
\usepackage{subfigure}

\begin{document}
\title{Spectral dimension as a probe of the ultraviolet continuum regime of causal dynamical triangulations}
\author{Thomas P. Sotiriou,$^{1,2}$ Matt Visser,$^3$ and Silke Weinfurtner,$^{1}$}
\affiliation{$^1$ SISSA - ISAS, Via Bonomea 265, 34136, Trieste, Italy {\rm and} INFN, Sezione di Trieste\\ 
$^2$ DAMTP, CMS, University of Cambridge, Wilberforce Road, Cambridge, CB3 0WA, UK\\
$^3$ MSOR, Victoria University of Wellington, PO Box 600, Wellington 6140, New Zealand}
\begin{abstract}
We explore the ultraviolet continuum regime of causal dynamical triangulations, as probed by the flow of the spectral dimension. We set up a framework in which one can find continuum theories that can in principle fully reproduce the behaviour of the latter in this regime. In particular, we show that, in $2+1$ dimensions, Ho\v rava--Lifshitz gravity can mimic the flow of the spectral dimension in causal dynamical triangulations to high accuracy and over a wide range of scales. This seems to provide evidence for an important connection between the two theories.
\end{abstract}
\maketitle
The first serious effort to implement discretized geometries into the framework of general relativity dates back to 1961 and the development of Regge calculus~\cite{Regge:1961lr}. Since then there has been persistent interest  in a variety of discrete quantum gravity models~\cite{Williams:2006kp}. One particularly interesting variant is causal dynamical triangulations (CDT) \cite{Ambjorn:1998xu}, in which a geometry emerges as the sum of all possible triangulations (modulo diffeomorphisms) obeying a global time foliation. The sum is evaluated using the path-integral formalism, such that every history is weighted using a variant of Regge calculus, where the edge lengths of the fundamental building blocks, the $d+1$ dimensional simplices, is kept fixed. The various different histories can be viewed as all possible fluctuations in the geometry, and they differ in the number of simplices and in the manner the latter are glued together.

In a discrete model, such as CDT, part of the challenge is to find suitable probes for the continuum limit. The latter has been investigated numerically, working at a fixed volume (fixed number of simplices) and using a Monte Carlo algorithm~\cite{Ambjorn:2004qm,
Benedetti:2009ge}. The {\em spectral dimension} has been proposed as a probe in CDT and has a longer history in lattice quantum gravity, see Refs.~\cite{Ambjorn:2005db, Horava:2009if} and references therein. This can be thought of as the effective dimension as probed by an appropriately defined (fictitious) diffusion process (random walk). More concretely, a diffusion process from point $\bf{x}$ to point ${\bf x}'$ in (fictitious) diffusion time $s$ is characterized by the probability density $\rho({\bf x},{\bf x}', s)$. The average return probability $P(s)$ is defined as $\rho({\bf x},{\bf x}, s)$ averaged over all points in space. The spectral dimension is then defined as
\begin{equation}
d_s=-2\frac{\mathrm{d} \ln P(s)}{\mathrm{d}\ln(\mathrm{s})}.
\end{equation}
In CDT the diffusion is a discretized, stochastic process. The spectral dimension is not constant, but instead changes with $s$, and consequently with the  length scale.

Clearly, the concept of the spectral dimension is not limited to CDT or discrete theories. One could consider a diffusion process on a manifold, as was proposed in Ref.~\cite{Horava:2009if}. Then the spectral dimension is defined same as above and  $\rho({\bf x},{\bf x}', s)$ is determined by the diffusion equation
\begin{equation}
\label{Eq:DiffusionProcess}
\frac{\partial \rho(\mathbf{x},\mathbf{x}',s)}{\partial s} + \hat{D} \rho(\mathbf{x},\mathbf{x}',s)  = 0.
\end{equation}
The choice of the differential operator $\hat{D}$ corresponds to the ``type" of diffusion process being considered. 

For example, when $\hat{D}$ is the 3-dimensional Laplacian and $s$ is identified with real time Eq.~(\ref{Eq:DiffusionProcess}) becomes the heat equation. Instead, when a diffusion process is to be used as a probe of geometry (or kinematics), $s$ becomes a fictitious diffusion time and ${\bf x}$ represents a point in spacetime (in analogy with what was mentioned above for CDT). A natural choice for $\hat{D}$ then seems to be the propagator of perturbations in spacetime.
For instance, one could have $\hat{D}=g^{\alpha\beta}\nabla_{\alpha}\nabla_{\beta}$, where $g_{\alpha\beta}$ is the (Lorentzian) metric on a manifold (Greek indices run from $0$  to $d$, which denotes the number of spatial dimensions) and we have performed a Wick rotation ($t\to -i t$). In this case, for large $s$ the corresponding spectral dimension will probe the geometry associated with $g_{\alpha\beta}$. For small $s$, given that spacetime is flat in a sufficiently small neighborhood of each point, the spectral dimension will actually probe the kinematics associated with $\hat{D}$ \cite{SVW1}. So, when $\hat{D}$ reduces to the flat d'Alembertian at small scales, then $d_s \to d+1$.   Note that if  $\hat{D}$ is a more complicated differential operator at small scales ({\em e.g.}~due to ultraviolet corrections) this will be encoded in the small $s$ behaviour of $d_s$.

Given the above, it is very tempting to use the spectral dimension as a probe of the continuum limit of discrete theories, and CDT in particular, or as a potential link to continuum theories. In fact, in Ref.~\cite{Benedetti:2009ge} the large $s$ behaviour of $d_s$ was matched to the outcome of a diffusion process in  a curved manifold with the geometry of a stretched sphere. However, the small $s$ (ultraviolet) behaviour of $d_s$ in CDT is, to date,
not well understood. 
This regime is believed to encode the influence of quantum corrections and is expected to provide the most prominent hint towards the effective field theory arising from CDT. Our goal is to demonstrate that one can indeed extract considerable information regarding the ultraviolet continuum regime of CDT from the small $s$ behaviour of $d_s$, and make a first crucial step towards identifying the characteristics of an (effective) continuum theory that could reproduce this behaviour.

The key idea is that $d_s(s)$ can be used to determine $\hat{D}$, which in turn characterizes (to some extent) a continuum theory. In Ref.~\cite{SVW1} we argue that in principle one can indeed determine the dispersion relation associated with $\hat{D}$ when $d_s(s)$ is known in closed form. However, when the latter is known only in tabulated form, as is the case for CDT, one should instead rely on the techique of non-linear regression to reconstruct the dispersion relation from $d_s(s)$. This requires some educated guess for the general form of the dispersion relation.

Consequently, what one can do in practice is to choose a continuum theory of gravity that shares some fundamental characteristic(s) with CDT and check if it can reproduce the behaviour of $d_s(s)$.
Clearly, this would not prove that the theory in question is the continuum limit of CDT: the spectral dimension does not carry all the information about the theory (as we will see in more detail below). Additionally, due to the differences in the definitions of the spectral dimension in the discretium and in the continuum there is still some ambiguity on the nature of the linkage it provides between discrete and continuum theories. Nonetheless, even with these caveats in mind, identifying a continuum theory that could fully reproduce the behaviour of the spectral dimension for small $s$ in CDT certainly seems to be a crucial step in understanding its ultraviolet continuum regime.

A suggestion for the candidate continuum theory has been made in Ref.~\cite{Horava:2009if}. It was noticed there that the preferred causal structure of CDT, imposed by a preferred foliation by slices of constant time, is reminiscent of the symmetries of Ho\v rava--Lifshitz (HL) gravity~\cite{Horava:2009uw}. (See Ref.~\cite{Sotiriou:2010wn} for a brief review including viability constraints.) Indeed, the latter is a theory with a preferred spacelike foliation, described by a scalar field. The existence of the preferred foliation allows for higher order spatial derivatives in the theory without having higher order time derivatives. This leads to significantly improved ultraviolet behaviour which actually renders the theory power-counting renormalizable \cite{Horava:2009uw,Visser:2009ul}, at the expense of giving up Lorentz invariance.

The action of HL gravity is, in the preferred foliation,
\begin{equation}
S=\frac{M_\mathrm{pl}^2}{2} \int \mathrm{d}^d x \, \mathrm{d}t \, N \sqrt{g} \left( K^{ij} K_{ij} - \lambda K^{2} + \mathcal{V} \right),
\end{equation}
where  $K_{ij}=\left( \dot{g}_{ij} - \nabla_i N_j - \nabla_j N_i \right)/(2N) $ is the extrinsic curvature, $N$ is the lapse, $N_j$ the shift and $g_{ij}$ is the induced metric on the spacelike hypersurfaces (Latin indices run from $1$ to $d$). $M_\mathrm{pl}$ is the Planck mass and $\lambda$ is a dimensionless running coupling. $\mathcal{V}$ is the part of the Lagrangian which contains only spatial derivatives. Power-counting renormalizability requires that it includes terms with at least $2d$ derivatives. Generally, $\mathcal{V}$ should include all terms compatible with the symmetries of the theory \cite{Sotiriou:2009gy,Blas:2009qj}. (For $\lambda=1$, $\mathcal{V}=R$, where $R$ is the Ricci scalar of $g_{ij}$, the theory reduces to general relativity.) 
 
 We will focus here on $2+1$ dimensions ($d=2$) mainly because, due to computational limitations, lower dimensional simulations are expected to produce more accurate data sets for our purposes. For $d=2$ we have \cite{Sotiriou:2011dr}
 \begin{eqnarray}
 {\mathcal V}&=&\xi R +\eta\, a^2 +g_1 \,R^2+g_2\, \nabla^2R+g_3\,a^4+g_4\, Ra^2\nonumber\\&&
+g_5 a^2 (\nabla^j a_j) + g_6 (\nabla^i a_i)^2 + g_7 (\nabla_i a_j) (\nabla^i a^j),\;
\end{eqnarray} 
 where $a_{i}=\partial_i \mathrm{ln} N$, $a^2=a_ia^i$ and $\eta$ and $\xi$ are dimensionless coupling, while the $g_{i}$ couplings have dimensions of an inverse mass squared. We have neglected the cosmological constant as it is irrelevant for our purposes.
 
We are interested in small length scales where curvature effects are negligible and quantum effects are expected to be important. So, for the diffusion process through which we define $d_s$ for HL gravity it is sufficient to use linearized propagators around flat space for $\hat{D}$. As we have shown in Ref.~\cite{Sotiriou:2011dr}, 
in $2+1$ dimensional HL gravity, the foliation-defining scalar with dispersion relation 
\begin{equation}
\label{Eq:DispRelMostGen2+1}
\omega^2= \frac{P_\mathrm{1}(k^2)}{P_\mathrm{2}(k^2)}=\mathtt{A}k^2 \frac{1+ \mathtt{B}k^2 + \mathtt{C}k^4}{1+\mathtt{D}k^2},
\end{equation}
is the only propagating mode. Here 
\begin{eqnarray}
\label{Eq:A} &&\mathtt{A}=\frac{1-\lambda}{1-2\lambda} \frac{\xi^2}{\eta}, \quad 
\mathtt{B}= \frac{2\left( 2 \eta \, g_1 + \xi \, g_2   \right)}{\xi^2}, \\
 &&\mathtt{C}=  \frac{g_2^2 - 4g_1(g_6+g_7)}{\xi^2},\quad \mathtt{D}= \frac{g_{6}+g_7}{2\eta}. \nonumber
\end{eqnarray}
 (The dispersion relation for the scalar mode is $3+1$ dimensional HL gravity is qualitatively the same \cite{Blas:2009qj}.)  

In Ref.~\cite{SVW1} we discuss in detail how one can define the spectral dimension associated with a general dispersion relation of the form $\omega^2=f(k^2)$. In this case the formal solution of Eq.~(\ref{Eq:DiffusionProcess}) after Wick rotation is 
\begin{equation}
\label{Eq:ProbabilityDensity}
\rho(\mathbf{x},\mathbf{x}',s)=\int{ \frac{\mathrm{d}^2 k \, \mathrm{d}  \omega}{(2\pi)^{3}} 
\mathrm{e}^{i(\mathbf{k}\cdot(\mathbf{x}-\mathbf{x}'))} \, \mathrm{e}^{-s(\omega^2 + f(k^2))} \, .}
\end{equation}
A straightforward calculation yields
\begin{equation}
\label{Eq:SpecDimHL}
d_s=1+ 2s \, \frac{\int{f(k^2) \, k^{d-1} \mathrm{e}^{-s\, f(k^2)} \; \mathrm{d} k }}{\int{ k^{d-1} \mathrm{e}^{-s\, f(k^2)} \; \mathrm{d} k }} \, .
\end{equation}
This formula is directly applicable to the dispersion relation in Eq.~(\ref{Eq:DispRelMostGen2+1}).

We have now laid out all of the technical tools that allow us to fit the spectral dimension for CDT in the ultraviolet continuum regime  in $2+1$ dimensions using the simulations of Ref.~\cite{Benedetti:2009ge}. These simulations were repeated for $6$ different values of the number of simplices $N_s=[40, 50, 70, 100, 140, 200]\times 10^3$. For each independent simulation $N_s$ was held fixed, and one sums over $1000$ different histories (geometries). On each history the (fictitious) diffusion process was  implemented and used to calculate the corresponding spectral dimension $d_s$.

In Ref.~\cite{Horava:2009if}  the spectral dimensions of CDT and HL gravity were compared at the two limits of the ultraviolet continuum regime and they were found to be in good agreement. Both theories seem to predict that at small scales $d_s$ flows to 2, whereas at length scales large enough for the quantum correction to become largely subdominant, but small enough for curvature correction to be unimportant, $d_s$ flows to 3, the number of topological dimensions. Encouraging as it may be, this result is far from being conclusive, given that there is an infinity of curves joining two points.  

Our goal is far more ambitious: we numerically evaluate the spectral dimensions in Eq.~(\ref{Eq:SpecDimHL}) for a dispersion relation of the kind~(\ref{Eq:DispRelMostGen2+1}), and fit the CDT data using the method of non-linear regression. This will allow us to obtain the parameters of the dispersion relation that provide the best fit for the {\em whole} ultraviolet continuum region, and access how good such a fit can be. This will provide a conclusive answer to whether HL gravity can reproduce the behaviour of $d_s$ for CDT in this regime.

This procedure will not uniquely determine all of the parameters of HL gravity. First of all, it is already clear from Eqs.~(\ref{Eq:A})
that $g_3$, $g_4$ and $g_5$ do not enter into the linearized propagator and, therefore, they cannot be determined without taking into account curvature corrections. Secondly, there are $7$ couplings that do enter in the propagator, but only in $4$ combinations. This implies that any observable that probes the dispersion relation will give us some but not full insight into the fundamental theory. Finally, without loss of generality, we can choose to work in units where the infrared light speed is one, i.e.~$\mathtt{A}=1$. This amounts to the rescaling $\omega\to \omega/\sqrt{\mathtt{A}}$. 

%
\begin{figure*}
\centering

\subfigure[]{
   \includegraphics[width=5.59cm]{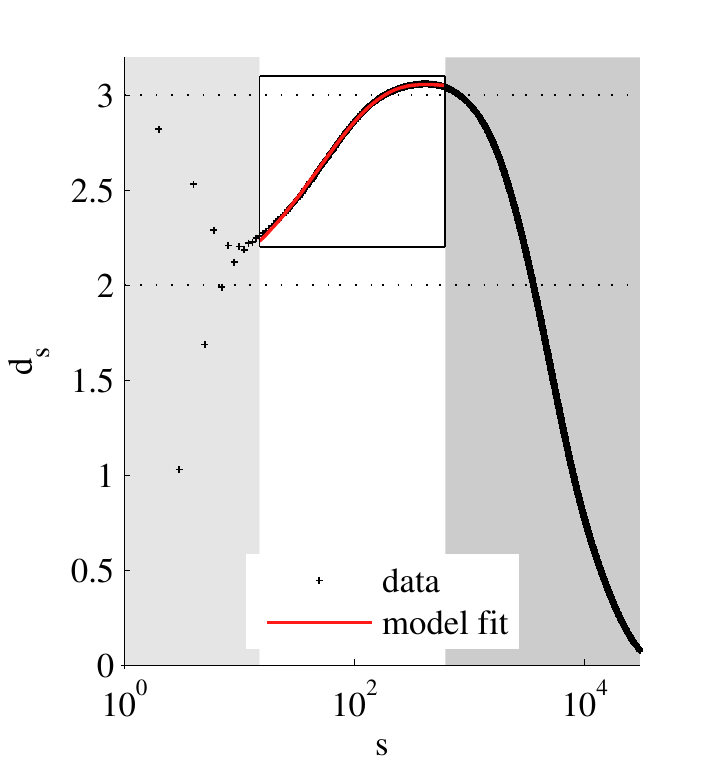}
   \label{fig:subfig1}
 }
\,
 \subfigure[]{
   \includegraphics[width=5.59cm]{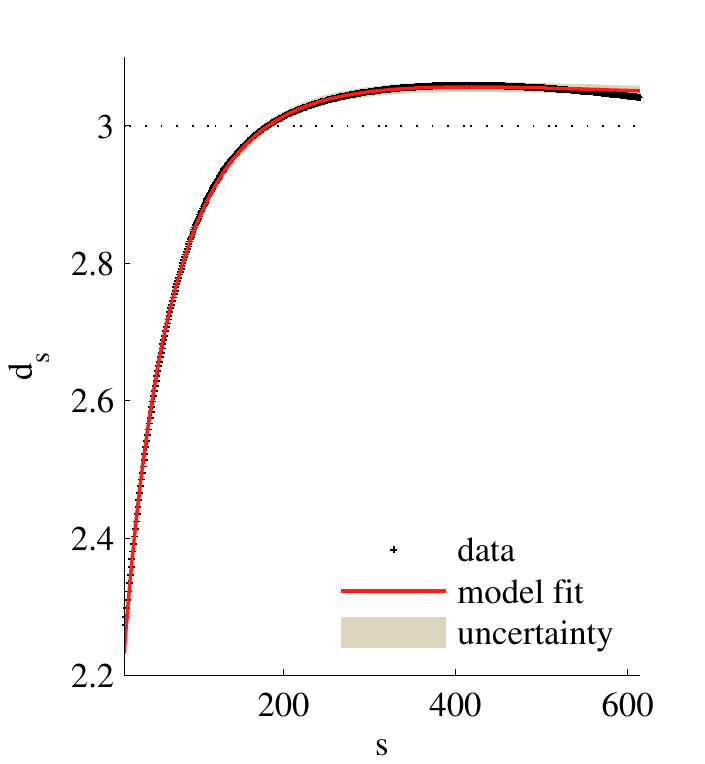}
   \label{fig:subfig2}
 }
\,
 \subfigure[]{
   \includegraphics[width=5.59cm]{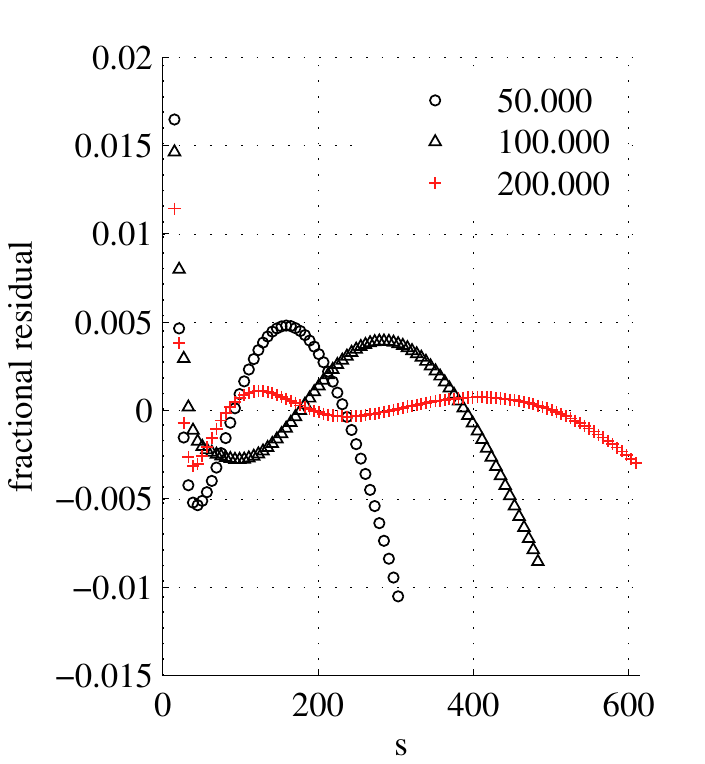}
   \label{fig:subfig3}
 }

\label{myfigure}
\caption{(a) The spectral dimension as a function of (fictitious) diffusion time $s$ (semilog scale). Crosses represent data points for CDT for simulation with $200,000$ simplices, whereas the red solid line represent the fit based of HL gravity. The region where discretium effects are important is lightly shaded. The region where curvature effects dominate is darkly shaded. The intermediate is the ultraviolet continuum region. (b) Zoom into the latter region. (c) Individual residuals for the model fits for $50,000$, $100,000$ and $200,000$ simplices. The residuals grow as expected at very small and very large scales under the influence of discretium and curvature effects respectively. The oscillatory pattern at intermediate scales seems to indicate some systematic deviation from the model, which however, tends to disappear as the  number of simplices is increased. For $200,000$ simplices the fractional residual remains below $0.5\%$ almost throughout the entire ultraviolet continuum region.}
\end{figure*}

In Fig.~\ref{fig:subfig1} we plot the spectral dimension $d_s$ as resulting from the CDT simulation with the largest number of simplices. 
As illustrated with the differently shaded areas, the diffusion process can be divided into three physically different regions. For very small $s\in[1 \, , \, 15]$ the spectral dimension shows discrete behaviour as expected. For very large $s\in[614 \, , \, 29999]$, the behaviour of the spectral dimension is expected to be dominated by curvature effects. This is the region were the flow of $d_s$ is compatible with a diffusion process on a continuous manifold with the geometry of a stretched sphere \cite{Benedetti:2009ge}. The intermediate region, where a continuous behaviour emerges but quantum corrections are pre-dominant, is the one we are interested in.  Our model-fit for this region ($s\in[15 \, , \, 614]$) is represented by a red solid line. Fig.~\ref{fig:subfig2} zooms into this region.

We have chosen to present a fit for the simulation with the larger number of simplices, because the higher the number of simplices, the larger the volume of the universe will be in the simulations. If the volume is not large enough, curvature effects can kick in before quantum effects become subdominant. There might then be no regime where both effects can be neglected, so no regime where the diffusion effectively takes place in flat space and the spectral dimension flows to $3$. This undesirable behaviour is indeed observed in some of the simulations with fewer simplices.

In Fig.~\ref{fig:subfig1} we see that for the $200,000$ simplices simulation $d_s$ starts at around $2$ in the ultraviolet continuum regime, it reaches $3$ in a regime where quantum effects are apparently still non-negligible and push $d_s$ slightly over $3$. Eventually, $d_s$ reaches $3$ for a second time and then drops to lower values under the influence of curvature corrections. Remarkably, our model can reproduce this behaviour very well in all the region where curvature corrections are negligible, including the part of the curve that exceeds $3$. The corresponding values for the parameters of the model, in units where $\mathtt{A}=1$, are $\mathtt{B}=-1.18\pm0.22$, $\mathtt{C}=344.47\pm3.11$, $\mathtt{D}=10.08\pm0.30$, which are compatible with technically natural values for the various couplings of HL gravity.

As an indication of how good our fit is, we present in Fig.~\ref{fig:subfig3} the individual residuals (difference between data and best fit)  for 3 simulations involving different number of simplices: $N_s=[50, 100, 200]\times 10^{3}$. For all $3$ cases the residuals grow for very small values of $s$. We attribute this to the fact that at some sufficiently small $s$ the discrete nature of CDT becomes important. Note that, even though we do not attempt to fit the lightly-shaded area of Fig.~\ref{fig:subfig1} where the data points violently oscillate, the absence of obvious oscillations in $d_s$ does not imply the complete absence of discreteness effects. 

Moving to intermediate values of $s$, a somewhat worrisome feature of the residuals is that they are not randomly scattered, but instead they exhibit some oscillatory pattern. This indicates that the fit might be missing a systematic effect. Nevertheless, the fact that the amplitude of the oscillations clearly reduces significantly as one moves to larger number of simplices is very encouraging, as it indicates that this effect is likely to become negligible as the size of the simulations increases. Additionally, this unknown effect might well be the subdominant contribution of the aforementioned discreteness effects.

The absolute magnitude of the residuals for all simulations becomes large again for large enough values of $s$. This is expected and signals the scale at which curvature corrections, which we neglect, become important. Consequently, the fact that the residuals start growing at a value of $s$ that is lower that the value for which $d_s=3$ for the second time (or even for the first time for lower numbers of simplices) seems to indicate that, even for the largest number of simplices, the universe is not large enough to allow  one to comfortably define a patch of spacetime that is large enough for quantum effects to be negligible, and at the same time small enough for curvature corrections to be largely subdominant.

It should be pointed out that we have neglected the running of couplings, even though $d_s$ is not necessarily a universal observable throughout the region considered. We are not fitting the deep UV, and so it is not unreasonable to assume that the running is not very significant. Our results {\em a posteriori} justify this expectation.

We have attempted to fit the CDT data with other types of dispersion relations, either inspired by lattice field theory, or just simple polynomials, which typically arise in projectable HL gravity where the lapse function is taken to be space independent \cite{Horava:2009uw,Sotiriou:2009gy}. However, the independent residual test largely favours rational dispersion relations, as in Eq.~(\ref{Eq:DispRelMostGen2+1}) \footnote{Projectable HL gravity type dispersion relations in $2+1$ dimensions could also be ruled out analytically as they are of the form $\omega^2=k^4$ \cite{Sotiriou:2011dr} and $d_s$ remains $2$ throughout.}. The flow of $d_s$ as defined in the discretium in CDT appears to match that of the foliation-defining scalar in HL gravity in the ultraviolet continuum regime, and not that of a test scalar field.

To summarize, we have shown that in $2+1$ dimensions HL gravity provides a very good fit (which is expected to  improve as simulations include a larger number of simplices) to the flow of the spectral dimension in CDT throughout the ultraviolet continuum regime. Even though this is by no means enough to argue that the former is the continuum limit of the latter, it does provide evidence for an important connection between the two theories. It also demonstrates beyond doubt, that the spectral dimension is a powerful tool for relating discrete and continuum theories and for gaining insight into their behaviour.

\begin{acknowledgments}
We are indebted  to D.~Benedetti, J.~Henson, and R.~Loll for providing the CDT data and for useful discussions. We also thank P.~Ho\v rava for enlightening comments and discussions. TPS and SW were supported by Marie Curie Fellowships. MV was supported by the Marsden Fund.
\end{acknowledgments}

\vskip-20pt


\end{document}